# Physics-informed neural networks for quantitative assessment of cancellous bone microstructure from photoacoustic signals

Shoukun Lyu[1], Haohan Sun[1], Shibo Nie[1], Weiya Xie[4], Ying Gu[1], Shiying Wu[5], Ya Gao[1,2,3*], Qian Cheng[1,2,3*]

[1]Institute of Acoustics, School of Physics Science and Engineering, Tongji University, Shanghai 200092, China.

[2]Department of Ultrasonography, Shanghai Disabled Persons' Federation Key Laboratory of Intelligent Rehabilitation Assistive Devices and Technologies, Yangzhi Rehabilitation Hospital (Shanghai Sunshine Rehabilitation Center), Tongji University School of Medicine, Shanghai 201619, China.

[3]National Key Laboratory of Autonomous Intelligent Unmanned Systems, Shanghai Research Institute for Intelligent Autonomous Systems, Tongji University, Shanghai 201210, China.

[4]College of Underwater Acoustic Engineering, Harbin Engineering University, Harbin 150001, China.

[5]School of Medical Imaging, Shanghai University of Medicine and Health Sciences, Shanghai 201318, China.

*Corresponding author E-mail: q.cheng@tongji.edu.cn, y.gao@ tongji.edu.cn

## Abstract

Artificial intelligence (AI) empowers innovative diagnostic tools for common diseases, yet its clinical application in skeletal health evaluation is constrained by unsatisfactory accuracy, owing to the inherent porous and poroelastic biophysical features of bone. To address such bottlenecks amid global population aging, this study targets skeletal health and develops a reliable AI framework for precise bone microstructural characterization. We proposed Biot-PINN, a physics-informed neural network embedded with Biot's poroelasticity theory to characterize mechanical responses and wave propagation in poroelastic bone tissues. By decoding photoacoustic signals encoding bone mineral and microstructural features, the framework enables automatic bone microstructural grading. Experimental results reveal that Biot-PINN reaches an accuracy of 97%, markedly surpassing traditional data-driven approaches and providing a robust solution for early skeletal health diagnosis.

## Introduction

The global burden of age-related osteoporosis and fractures is rising rapidly, creating an urgent clinical need for an accurate, early assessment of bone microstructure to guide intervention in high-risk individuals [1–3]. As a non-invasive, ionizing radiation-free, and molecule specific detection modality, the photoacoustic (PA) technique is well-positioned for high-precision skeletal health assessment. By exploiting the characteristic optical absorption of bone minerals, this technique enables direct visualization of molecular composition and microarchitectural features, supporting high-fidelity characterization of bone internal structure [4,5]. Nevertheless, robust quantification of bone microstructural parameters remains a major challenge, particularly for cancellous bone with its complex porous network [6,7]. Unlike homogeneous soft tissues, bone exhibits intricate fluid-solid coupling that cannot be fully characterized by conventional wave equations. The inherent wave propagation, accompanied by severe acoustic scattering and frequency-dependent attenuation, renders traditional numerical and analytical methods ineffective for precisely retrieving skeletal microstructural parameters through solving such highly complex inverse problems [8].

Given these limitations, artificial intelligence (AI) has emerged as a powerful alternative for medical signal analysis, as its strong representational capabilities enable automated extraction of intricate signal features to bypass the direct analytical solving of complex inverse problems [9–13]. Various machine learning architectures have been developed to decipher complex biological data, with convolutional neural networks (CNNs) excelling at spatial feature extraction [14–20], recurrent neural networks (RNNs) capturing temporal dynamics [21,22], and Transformers exploiting global long-range dependencies [23,24]. However, for complex biological media such as bone, the clinical translation of AI-based decoding for skeletal assessment remains limited by insufficient predictive accuracy [25]. This limitation arises not from stagnation in neural architecture, but from a fundamental methodological constraint, i.e., current deep learning models operate predominantly as statistical engines, excelling at detecting empirical correlations in large-scale data, yet lacking intrinsic mechanisms to encode domain-specific physical laws. When applied to complex, heterogeneous physical media such as bone, where mechanical, compositional, and microarchitectural properties interact through multiscale, bidirectional solid-fluid couplings, such purely data-driven approaches frequently yield predictions that violate fundamental biomechanical and biophysical principles [26–28]. To bridge this gap, next-generation architectures must integrate physics-informed priors directly into model design, embedding explicit constraints derived from continuum mechanics, poroelasticity, or tissue-level structure-function relationships. Such physics-integrated frameworks do not merely improve numerical accuracy; they enhance interpretability, ensure thermodynamic and mechanical consistency, and establish the scientific rigor and clinical accountability required for regulatory acceptance and real-world deployment in skeletal diagnostics.

To address these challenges, Physics-Informed Neural Networks (PINNs) offer a rigorous, equation-constrained learning paradigm that intrinsically embeds first-principles physical laws, rather than merely approximating them, into deep neural architectures. Initially formalized by Raissi et al. [29], PINNs exploit automatic differentiation to encode governing equations as physical residuals within the loss function. This constraint enforces strict adherence to physical conservation laws throughout training, thereby eliminating unphysical predictions that violate fundamental mechanics [25]. As a mesh-free, data-efficient framework, PINNs have demonstrated broad applicability across diverse domains, ranging from signal analysis [30–33] and physical field reconstruction [34,35] to material structure modeling [36,37]. Crucially, the efficacy of a physics-informed model is inherently governed by the fidelity of its underlying physical priors [25]. Therefore, integrating a biophysically grounded, bone-specific physical constitutive theory into the PINN architecture is essential to enable high-fidelity, physically interpretable inversion of microstructural parameters with clinical utility.

In this work, we propose Biot-PINN: a physics-informed, differentiable inverse framework for quantitative, noninvasive assessment of bone microstructure. Grounded in Biot's poroelasticity theory which rigorously couples solid matrix deformation and pore-fluid pressure diffusion in saturated porous media, we explicitly embed Biot's governing equations as constraints in the loss function. This enables end-to-end, gradient-based inversion of key microstructural parameters from broadband photoacoustic signals that encode both molecular mineral composition and hierarchical structural features [5,38,39]. By enforcing physical consistency at each optimization step, the proposed approach ensures the solution process complies with the acoustic characteristics of porous media, while effectively mitigating overfitting to noise or limited measurement data. To validate the proposed physics-informed approach, we perform numerical simulations and ex vivo experimental tests on bone models with varying porosity levels, mimicking the pathological progression from a healthy status to osteoporosis. The results demonstrate that embedding physical constraints can effectively suppress non-physical solutions and achieve superior accuracy relative to conventional purely data-driven models.

## RESULTS

### Framework overview and flowchart

We developed an AI-assisted framework for the quantitative characterization of bone microstructure, with the overall schematic illustrated in Figure 1.

In the data acquisition stage (shown in Figure 1b), we employ PA technology to capture bone microstructural information. Pulsed laser excitation induces thermoelastic expansion in bone tissue, generating acoustic waves that propagate through the trabecular network. Wave propagation is governed by poroelastic interactions between the solid bone matrix and interstitial fluid, as described by Biot's governing equations. The acquired multi-channel time-domain PA signals serve as inputs for the subsequent PINN-based quantitative inversion.

We focus on bone porosity ($\varphi$) as the key microstructural indicator of skeletal mechanical strength and fracture risk. In the bone porosity quantitative inversion stage (Fig. 1c), we develop a Biot-informed Physics-Informed Neural Network (Biot-PINN). This framework is engineered to achieve a deep integration of data-driven feature extraction with the rigorous physical constraints dictated by Biot's poroelasticity theory. We adopt a U-Net as the core architecture [40], a convolutional framework that has proven highly effective in capturing both temporal evolution and spatial correlations within multichannel medical signals. The network takes 128-channel time-domain PA signals as input, captures both the dynamic temporal evolution and inter-channel spatial correlations through its encoder-decoder pathway, and outputs the target parameter porosity ($\varphi$) along with two auxiliary physical fields representing the solid phase displacement ($u$) and the fluid phase displacement ($w$). Building upon these three network outputs, we construct a loss function comprising three complementary components. First, the supervised data loss ($\mathcal{L}_{data}$) directly compares the predicted porosity ($\varphi$) with ground truth labels using cross-entropy loss. Second, leveraging automatic differentiation [41], we compute the spatiotemporal derivatives of the predicted displacement fields (u, w) and substitute them, along with the predicted porosity, into Biot's governing equations to obtain physics residuals. These residuals yield two physics losses ($\mathcal{L}_{PDE_1}$, $\mathcal{L}_{PDE_2}$) enforcing momentum balance for the solid and fluid phases, respectively. Third, we introduce a reconstruction loss ($\mathcal{L}_{recon}$). We first compute the predicted pressure field from all three network outputs via Biot's constitutive relations, then quantify its discrepancy with the true pressure field reflected in the input PA signals. These three loss components are combined with weighting factors $\alpha$ and $\beta$ to form the total loss ($\mathcal{L}_{total}$), which guides the network training [42].

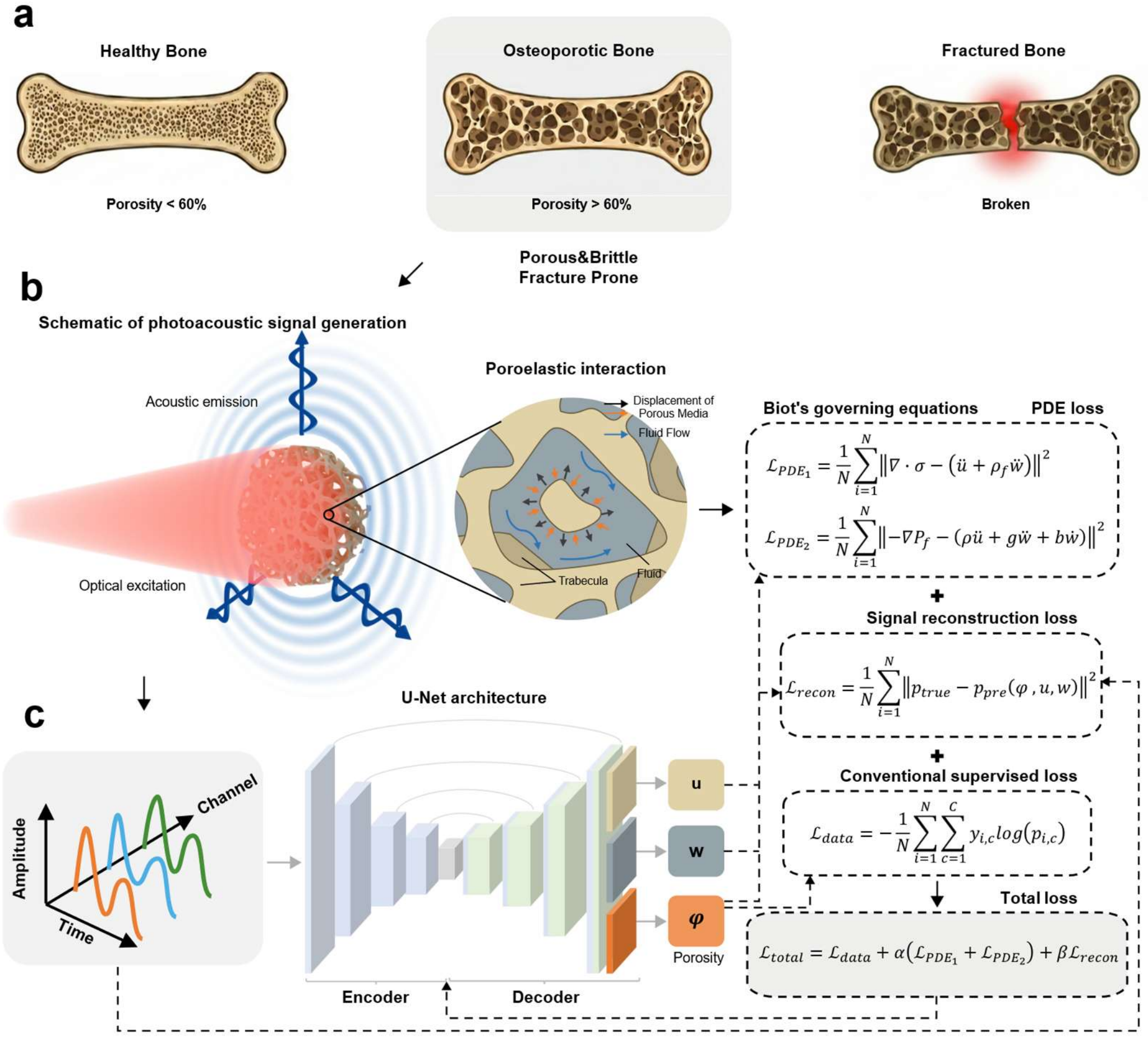

**Fig. 1 Schematic of Biot-PINN for bone microstructure assessment. a** Clinical motivation showing bone microstructure evolution from healthy (porosity < 60%) to osteoporotic (porosity ≥ 60%) and fractured states. **b** Photoacoustic signal generation mechanism: pulsed laser excitation induces thermal expansion in bone tissue, generating acoustic emission. The resulting signals carry the signature of microscopic poroelastic interactions within the trabecular bone—a biphasic process characterized by dynamic solid-fluid interaction. In this coupled system, the rapid expansion of the solid trabecular matrix and the motion of the interstitial fluid are mutually coupled, as illustrated by the concurrent local displacements (orange arrows) and fluid flow (blue arrows) within the pore network. These solid-fluid coupling dynamics are governed by Biot's equations, which describe the momentum balance for both phases. Accordingly, the signal encodes information about poroelastic interactions in trabecular bone, governed by Biot's equations that describe solid-fluid coupling dynamics. The physics-informed loss comprises two components ($\mathcal{L}_{PDE_1}$, $\mathcal{L}_{PDE_2}$) derived from the momentum balance equations for solid and fluid phases, respectively. **c Biot-**PINN-based quantitative inversion stage: U-Net architecture takes multi-channel time-domain signals as input and outputs target porosity ($\varphi$) along with auxiliary displacement fields ($u$, $w$). The total loss function ($\mathcal{L}_{total}$) combines three components: supervised data loss ($\mathcal{L}_{data}$) for porosity ($\varphi$), physics losses ($\mathcal{L}_{PDE_1}$, $\mathcal{L}_{PDE_2}$) derived from Biot's equations, and signal reconstruction loss ($\mathcal{L}_{recon}$) enforcing consistency between predicted pressure field and input signal, with weighting factors α and β balancing these constraints.

## Numerical Simulation Validation

To validate the effectiveness of the proposed Biot-PINN, we designed a gradient-porosity bone assessment simulation framework, with porosity levels ranging from 35% to 95% in 10% increments. The numerical simulation dataset was constructed based on Micro-CT scan data across the seven porosity levels, as shown by the representative samples visualized in Fig. 2a. This porosity range covers the microstructural variability of human bone tissue from physiologically healthy states (35%-55%) to pathological osteoporotic states (65%-95%) [43,44]. In simulations of PA signal generation and propagation, we used the elastic wave model provided by the open-source k-Wave MATLAB toolbox (Fig. 2b) [45]. The acoustic properties of the simulation media, including density, speed of sound, and attenuation coefficients for both compression and shear waves, are detailed in Table 1 [46–49]. This simulation generated a numerical dataset comprising 280 sample groups (40 samples per porosity level), with each signal containing 128 channels and 4096 temporal sampling points within a window of 4.73–26.22 μs (Fig. 2c). The dataset was partitioned into training, validation, and test sets at an 8:1:1 ratio.

Using this dataset, we evaluated the performance of the proposed Biot-PINN framework in porosity parameter prediction. To rigorously validate the effectiveness of incorporating physical constraints, we compared Biot-PINN against two purely data-driven baseline models. The first baseline was a U-Net architecture retaining only the data loss term, maintaining identical network structure and parameter count to PINN. The second was a Transformer architecture, which currently represents a modern and advanced architecture for processing complex sequence dependencies and medical diagnostic signals [23,24]. Figure 3a illustrates the training dynamics of the three models over 200 epochs. The loss curves (left panel) demonstrate that Biot-PINN achieved the lowest final loss, followed by U-Net, while Transformer plateaued at a substantially higher loss. The accuracy curves (right panel) show that Biot-PINN reached the highest final accuracy, outperforming both U-Net and Transformer. To provide a detailed breakdown of classification performance, Figure 3b presents the confusion matrices of the three models on the test set. The Biot-PINN model achieved 96% test accuracy, with perfect (100%) identification for six out of seven porosity classes (45%-95%), and only minor misclassification in the 35% class (75% correct, 25% misclassified as 45%). The U-Net model achieved 82% test accuracy, with the poorest performance in the 35% class (25% correct, 50% misclassified as 45%, 25% as 55%), 75% accuracy for the 45% and 55% classes, and perfect identification for high porosity classes (65%-95%). The Transformer model achieved 79% accuracy, with zero correct classification in the 55% class (75% misclassified as 45%, 25% as 65%), 75% accuracy for the 35% and 75% classes, and 100% for the remaining classes. Notably, the Biot-PINN model exhibited misclassifications only in the lowest porosity class, whereas both purely data-driven models showed classification errors across multiple low-to-medium porosity classes.

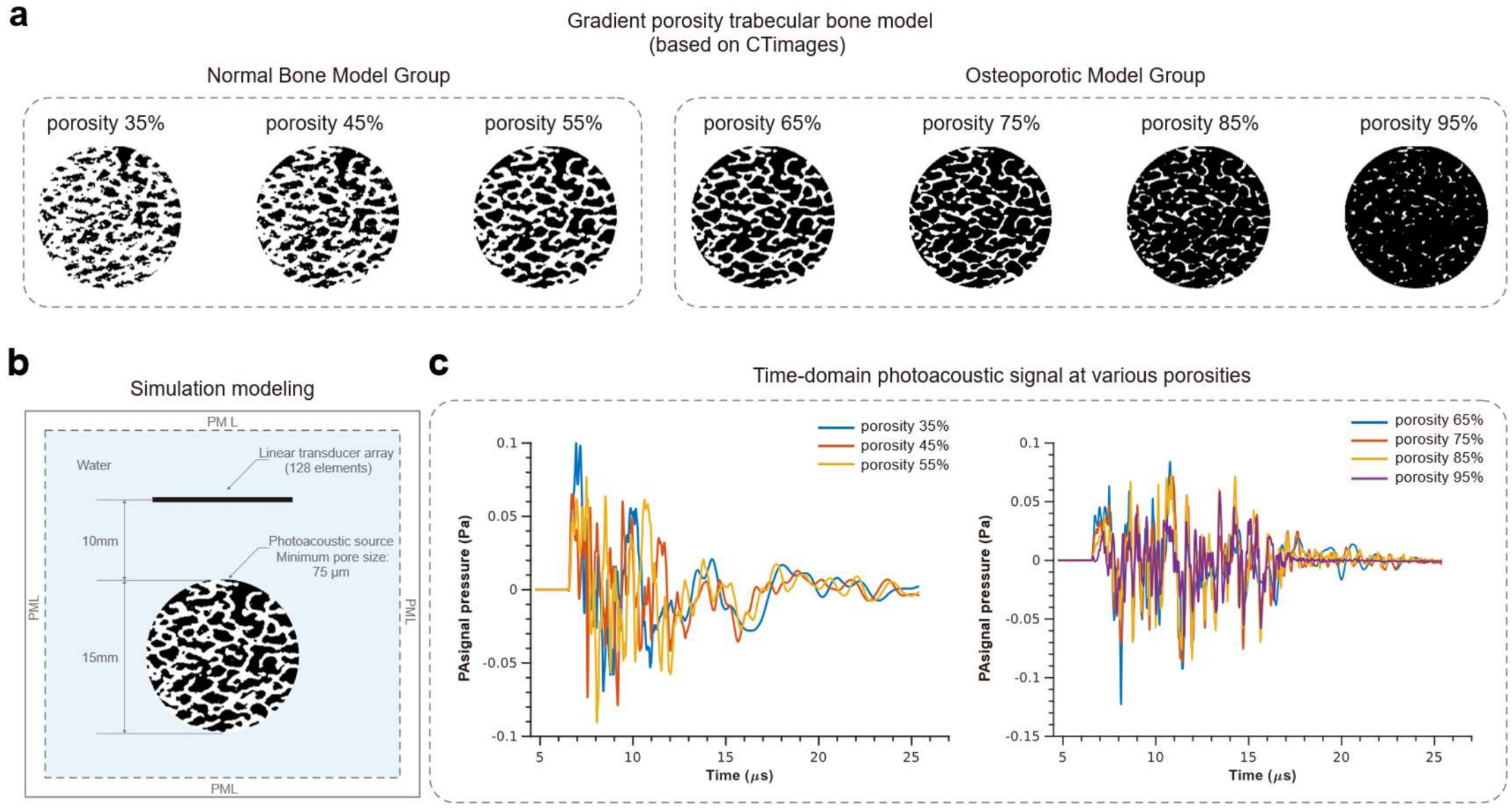

**Figure 2. Numerical simulation dataset generation and representative photoacoustic signals. a** Gradient porosity trabecular bone models based on CT images, spanning seven porosity levels (35%-95%). Models are divided into Normal Bone Model Group (porosity 35%-55%) and Osteoporotic Model Group (porosity 65%-95%). **b** Numerical simulation setup. The trabecular bone model (15 mm diameter, minimum pore size of 75 μm), serving as the photoacoustic source, is immersed in a water medium enclosed by perfectly matched layer (PML) boundaries. Acoustic waves are detected by a 128-element linear transducer array positioned 10 mm above the source. **c** Time-domain photoacoustic signals at various porosities. Left and right panels show representative multi-channel signals for seven porosity levels, demonstrating distinct amplitude and waveform variations across different porosity values.

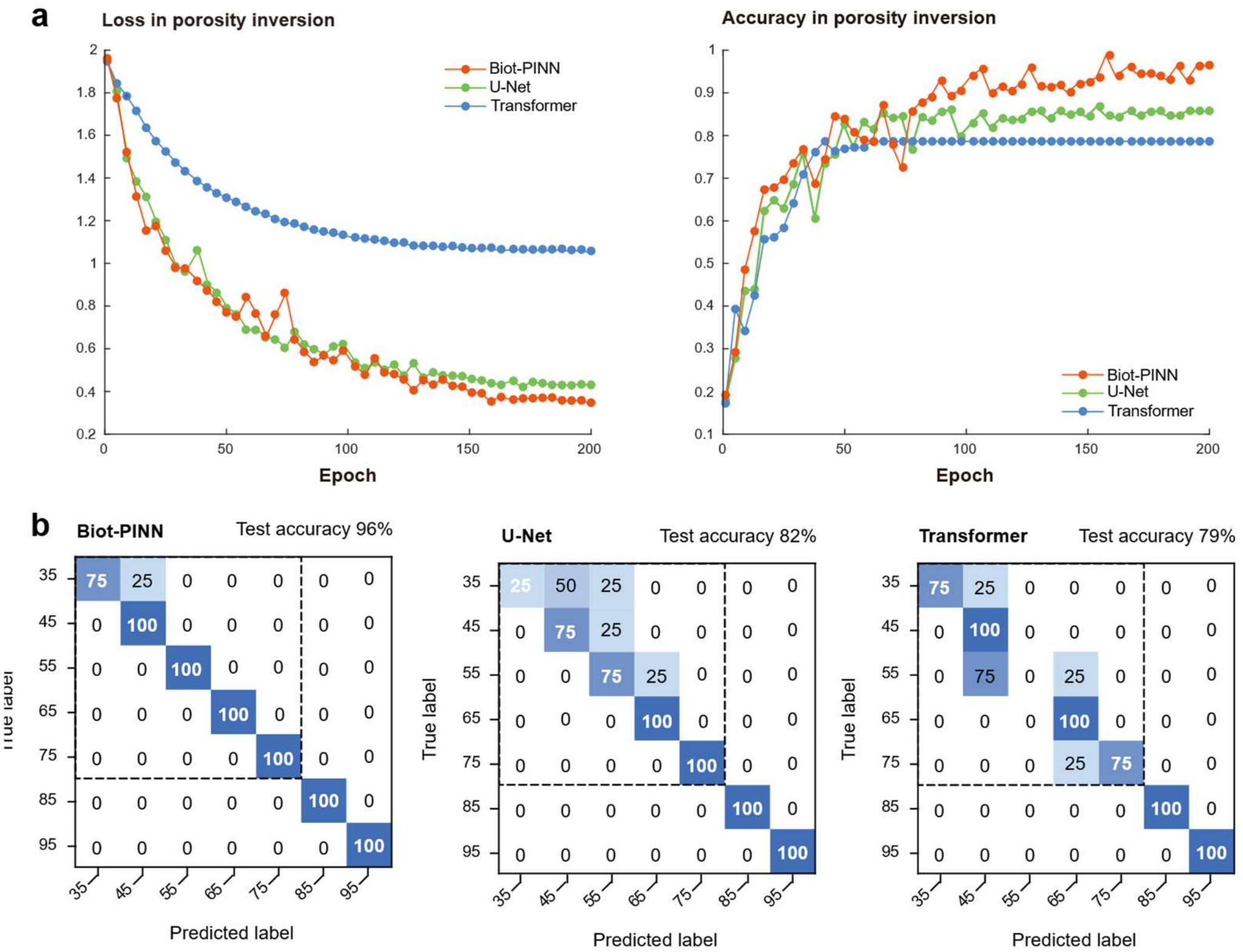

**Figure 3. Training dynamics and classification performance comparison of Biot-PINN and baseline models on simulation dataset. a** Training curves over 200 epochs for Biot-PINN (orange), U-Net (green), and Transformer (blue). Left panel: loss curves showing Biot-PINN achieved the lowest final loss, followed by U-Net, while Transformer plateaued at a substantially higher loss. Right panel: accuracy curves demonstrating Biot-PINN reached the highest final accuracy, outperforming both U-Net and Transformer. **b** Confusion matrices on the test set across seven porosity levels (35%-95%). Biot-PINN achieved 96% test accuracy with errors only in the lowest porosity class, while purely data-driven models (U-Net: 82%, Transformer: 79%) exhibited errors across multiple low-to-medium porosity classes. Color intensity represents the percentage of samples, with diagonal elements indicating correct classifications.

**Table1** Material parameters using in simulations of PA signal generation and propagation.

| | | Trabeculae | Water |
|---|---|---|---|
| Density(g/cm$^3$) | | 1800 | 1000 |
| Compression wave | SOS (m/s) | 2000 | 1500 |
| | α coefficient | 10 | 0.02 |
| Shear wave | SOS (m/s) | 1500 | 0 |
| | α coefficient | 17 | 0 |

**Ex Vivo experimental Validation**

To further validate the effectiveness of the proposed Biot-PINN and assess its clinical feasibility, we designed a gradient-porosity ex vivo photoacoustic bone assessment study using bovine bone as a surrogate for human bone tissue [50]. We prepared 23 bone slice samples with varying porosities, representing four stages of bone degeneration: healthy state (porosity <60%, n=5), early-stage degeneration (porosity 60-70%, n=6), mid-stage degeneration (porosity 70-80%, n=6), and advanced degeneration (porosity >80%, n=6). All samples underwent micro-CT scanning to establish ground-truth porosity measurements. We conducted photoacoustic detection using a Verasonics ultrasound system (Fig. 4a). To maximize data utility from limited samples, we acquired photoacoustic signals at eight rotational angles for each bone slice using a linear array transducer. Given the spatial heterogeneity of pore distribution within bone tissue, signals acquired at different angles capture distinct local pore distributions despite sharing identical overall porosity, thereby allowing each angular acquisition to be treated as an independent sample for model training. Each acquisition comprised 128 channels and 2688 time-domain sampling points. Figure 4b presents representative micro-CT images from each porosity category alongside their corresponding photoacoustic signals within the valid temporal window (8-30 μs). Ground-truth labels were determined by micro-CT measurements, while predicted labels were obtained from the proposed Biot-PINN using photoacoustic signals alone. For dataset partitioning, we employed a rigorous sample-level split strategy: two bone slices from each category were reserved as a independent test set, ensuring that all rotational acquisitions from these samples were excluded from the training pipeline. The remaining samples were divided into training, validation, and internal test sets at an 8:1:1 ratio. Following model training, the learned weights were applied to classify the eight completely unseen bone slices, providing an unbiased assessment of generalization performance on novel samples.

Using the ex vivo dataset, we evaluated the performance of the proposed Biot-PINN framework on real bone tissue classification. Consistent with the simulation study, we compared Biot-PINN against two purely data-driven baseline models: a U-Net architecture with identical network structure but only the data loss term, and a Transformer-based model. Figure 4c presents the confusion matrices of the three models on the independent test set of eight unseen bone samples. Biot-PINN achieved 97% test accuracy, with perfect (100%) classification for mid-stage and advanced degeneration samples, 94% accuracy for both healthy (6% misclassified as advanced) and early-stage classes (6% misclassified as healthy). The U-Net model achieved 91% test accuracy, with perfect identification for early-stage (100%), mid-stage (100%), and advanced (100%) degeneration, but struggled with the healthy class (63% correct, 37% misclassified as advanced). Notably, the PINN model achieved 94% accuracy on the healthy class, outperforming the U-Net model (63% accuracy) by 31%. The Transformer model achieved 66% accuracy, showing the poorest performance among the three models.

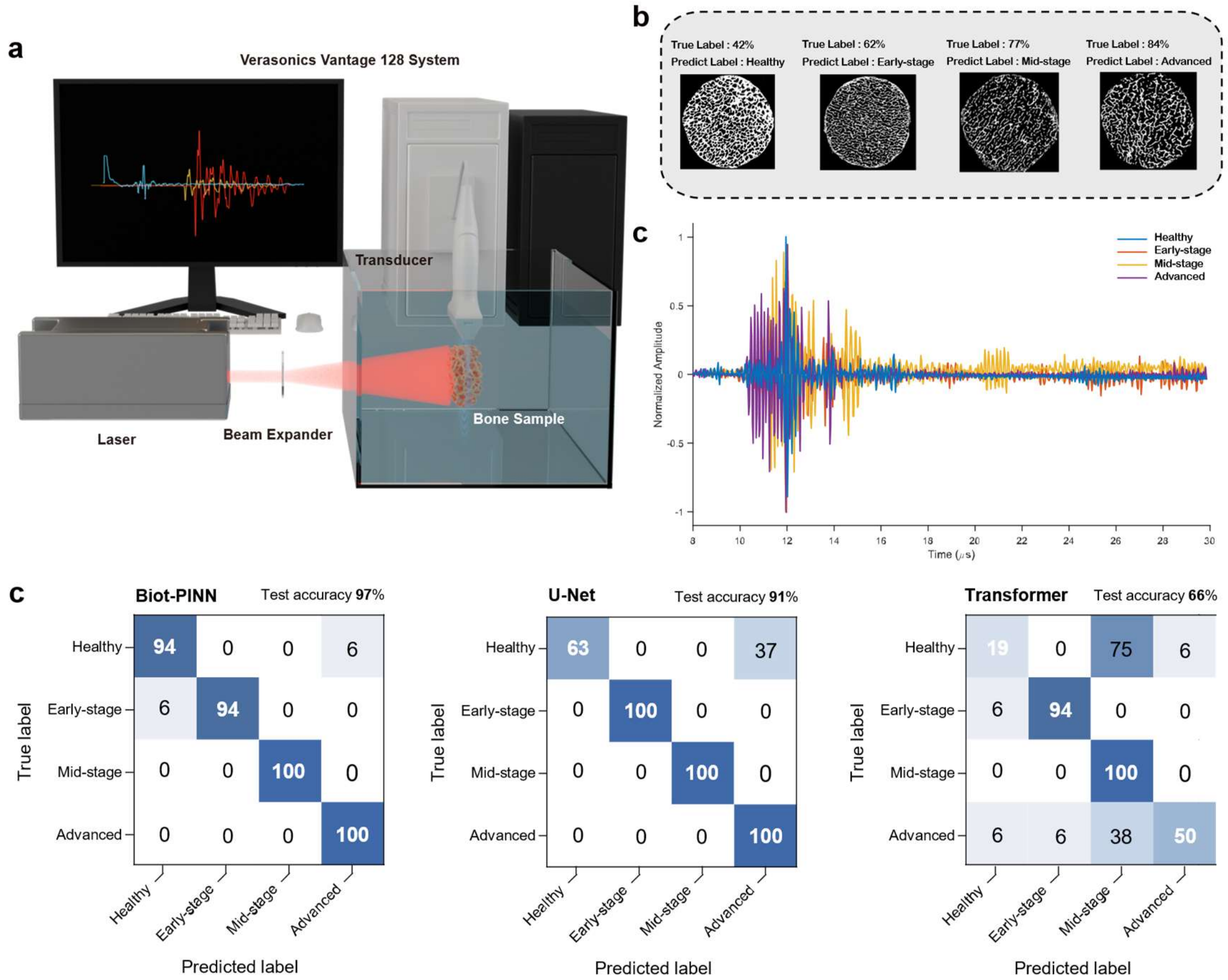

**Figure 4. Ex vivo photoacoustic detection system and classification performance comparison of Biot-PINN and baseline models. a** Schematic diagram of the experimental setup, comprising a laser source, beam expander, transducer, bone sample holder, and Verasonics Vantage 128 system for PA signal from cylindrical bone specimens. **b** Representative photoacoustic time-domain signals and corresponding CT cross-sectional images for four bone sample categories. CT images provide the ground truth labels (Healthy: 42% porosity, Early-stage: 62% porosity, Mid-stage: 77% porosity, Advanced: 86% porosity), while the predicted labels are determined by Biot-PINN based on the photoacoustic signals **c** Confusion matrices comparing classification performance across Biot-PINN (97%

accuracy), U-Net (91% accuracy), and Transformer (85% accuracy) on four porosity categories: Healthy (<60%), Early-stage (60-70%), Mid-stage (70-80%), and Advanced (>80%).

## DISCUSSION

In this study, we developed Biot-PINN to enable automated bone microstructural grading through the precise decoding of bone parameters from photoacoustic signals capturing molecular mineral and structural information. Through the successful integration of Biot's poroelasticity theory into the physics-informed neural network architecture, the model ensures that the inversion process remains strictly consistent with underlying physical laws to effectively suppress non-physical solutions. Crucially, by grounding neural learning in validated physical principles, Biot-PINN provides the high-fidelity accuracy and physical consistency essential for reliable clinical diagnostics. These findings establish a trustable and viable pathway for the clinical translation of AI-driven tools in bone health assessment.

The key advantage of our Biot-PINN over purely data-driven models lies in its multi-field joint learning mechanism [51,52]. Unlike conventional data-driven methods, which predict only porosity $\varphi$, our Biot-PINN simultaneously outputs three physical fields: $\varphi$, solid-phase displacement $u$, and fluid-phase displacement $w$, which are coupled through Biot equations. When porosity prediction deviates, this error propagates through the coupling relationships to manifest as multi-field residuals, increasing the physics loss. This mechanism enhances the model's sensitivity to porosity, improving inversion accuracy.

We observed that all three models are prone to classification errors in low-porosity regions. This stems from reduced signal discriminability due to weakened fluid-solid coupling [53]. Within the governing Biot equations used as physical constraints, the fluid phase displacement w incorporates a first-order time derivative that is absent in the solid phase displacement u, which causes the fluid dynamics to dominate the physical constraints. In low-porosity bone tissue, the restricted fluid volume limits this movement and results in diminished coupling effects. Conversely, although high-porosity regions contain a sparser solid matrix, the larger fluid volume and its dominant role in the equations facilitate more extensive fluid motion to produce relatively stronger fluid-solid coupling. This robust coupling excites fast and slow longitudinal waves with distinct velocities, providing the phase differences and energy ratios necessary for accurate classification [54]. In low-porosity regions, coupling degradation renders these features nearly imperceptible, reducing classification accuracy [55]. While Biot-PINN also faces challenges in weakly coupled low-porosity regions, its ability to explicitly model these interactions through the Biot equations allows it to maintain a lower error rate than purely data-driven models by leveraging prior physical knowledge. Future work could improve performance in these challenging regions through expanded training datasets or targeted data augmentation strategies.

The ex vivo experimental analysis revealed that certain healthy bone samples were misclassified as being at high risk for fracture, representing a diagnostic error that spans multiple clinical stages. These misclassifications primarily occurred because the experimental protocol utilized random detection angles to simulate the spatial variability and diverse probe orientations encountered in actual clinical settings. Given the inherent spatial heterogeneity of bone tissue, signals acquired from localized regions with larger pores can produce acoustic signatures that mimic the characteristics of advanced degeneration and thereby mislead the models. Notably, Biot-PINN significantly mitigated this challenge by incorporating physics-based constraints to achieve 94% accuracy on the healthy class compared to 63% for the U-Net model. This 31% improvement demonstrates that physics-informed priors effectively compensate for interference from localized structural anomalies that would otherwise dominate purely data-driven learning. While the U-Net remains susceptible to these non-uniform pore distributions, the integration of physical knowledge in Biot-PINN ensures greater robustness when confronting the heterogeneous tissue structures and varied detection angles typical of clinical practice.

To further address the challenges posed by spatial heterogeneity and non-uniform pore distribution [56,57], future research could expand upon the current homogenization assumption by incorporating auxiliary microstructural parameters such as trabecular thickness and connectivity. While the current Biot-PINN model effectively captures average porosity, a multiscale modeling approach could integrate classical Biot equations for macroscale efficiency with finer geometric descriptions at the microscale. These refinements would enhance diagnostic performance through more granular physics without altering the fundamental physics-informed framework. Nevertheless, classical Biot equations remain capable of accurately characterizing acoustic propagation in bone at the macroscale. Our simulation and ex vivo experimental results confirm that Biot-PINN demonstrates robust performance even in heterogeneous real bone samples. Thus, porosity, despite losing microscale detail, effectively characterizes overall bone mechanical and acoustic properties, supporting clinical bone quality assessment.

The transition from laboratory settings to clinical practice involves adapting optical delivery strategies to account for the presence of overlying soft tissues [4]. While our current ex vivo experiments utilize uniform illumination of bone cross-sections to ensure consistent signal generation, clinical applications would likely employ ring-shaped illumination to facilitate multi-angle energy deposition into deeper tissues [58]. To accommodate the light attenuation and scattering that occur as energy propagates through layered skin and muscle, future iterations of the Biot-PINN can be enhanced by incorporating specific attenuation terms into the physics-informed loss functions. Such refinements will advance clinical translation while preserving the established physics-informed foundation. Such refinements will advance clinical translation while preserving the established physics-informed foundation.

## METHODS

### Biot's theory of poroelasticity

Trabecular bone is a complex fluid-saturated biphasic porous medium. Its structural architecture is characterized by an interconnected solid frame of trabeculae, with its internal pore space filled with fluid-like bone marrow. This typical poroelastic nature dictates that its mechanical behavior cannot be accurately described by simplified single-phase models. In this context, the Biot-Willis theory provides a rigorous theoretical framework to comprehensively characterize the elastic, inertial, and viscous coupling effects between the solid and fluid phases. These coupling effects are well-suited for describing the complex wave propagation within the bone microstructure.

In this framework, we denote the solid phase displacement as $u$ and the fluid phase displacement as $u_f$. To describe the fluid motion relative to the solid skeleton, we introduce the relative fluid displacement w:

$$w = \varphi\left(u_f - u\right) \quad (1)$$

where $\varphi$ is the porosity, defined as the ratio of pore volume to total volume.

According to Biot's theory, the coupled equations of motion governing wave propagation in porous media can be expressed as:

$$\nabla \cdot \sigma = \rho \frac{\partial^2 u}{\partial t^2} + \rho_f \frac{\partial^2 w}{\partial t^2} \quad (2)$$

$$-\nabla P_f = \rho \frac{\partial^2 u}{\partial t^2} + g \frac{\partial^2 w}{\partial t^2} + b \frac{\partial w}{\partial t} \quad (3)$$

In the above equations, $\sigma$ is the total stress tensor of the porous medium, and $P_f$ is the pore fluid pressure. The bulk density is $\rho = (1 - \varphi)\rho_s + \varphi\rho_f$, where $\rho_s$ and $\rho_f$ represent the densities of the solid skeleton and pore fluid, respectively. The parameters $g = S\rho_f/\varphi$ and $b = \eta_f/\kappa_m$ denote the fluid-phase mass and viscous coupling coefficients, respectively, where $S$ is the structure factor and $\kappa_m$ is the permeability.

**Physics-informed neural network**

Based on the aforementioned Biot equations, the optimization process of the PINN needs to adhere to the PDE loss specified by Eq. (2) and Eq. (3), which is calculated based on the three outputs of the network.

$$\mathcal{L}_{PDE_1} = \frac{1}{N}\sum_{i=1}^{N}\left\|\nabla \cdot \sigma - \left(\rho\ddot{u} + \rho_f\ddot{w}\right)\right\|^2 \quad (4)$$

$$\mathcal{L}_{PDE_2} = \frac{1}{N}\sum_{i=1}^{N}\left\|-\nabla P_f - \left(\rho\ddot{u} + g\ddot{w} + b\dot{w}\right)\right\|^2 \quad (5)$$

where superscript $i$ denotes the $i$th sample and N denotes the number of samples, the double dot notation denotes the second time derivative and the single dot denotes the first time derivative. To ensure that the predicted reconstruction field is consistent with the ground-truth field reflected by the actual signals, a reconstruction loss is incorporated into the objective function:

$$\mathcal{L}_{recon} = \frac{1}{N}\sum_{i=1}^{N}\left\|p_{true} - p_{pre}(\varphi, u, w)\right\|^2 \quad (6)$$

where $p_{true}$ represents the input photoacoustic signal and $p_{pre}(\varphi, u, w)$ is the predicted pressure field reconstructed from the three network outputs. The data loss is formulated using cross-entropy:

$$\mathcal{L}_{data} = -\frac{1}{N}\sum_{i=1}^{N}\sum_{c=1}^{C} y_{i,c}\, log\left(p_{i,c}\right) \quad (7)$$

where $C$ represents the number of categories for the porosity classification, $y_{i,c}$ is the binary indicator (0 or 1) if class label $c$ is the correct classification for sample $i$, and $p_{i,c}$ is the predicted probability that sample $i$ belongs to class $c$. The total function is formulated as:

$$\mathcal{L}_{total} = \mathcal{L}_{data} + \alpha\left(\mathcal{L}_{PDE_1} + \mathcal{L}_{PDE_2}\right) + \beta\mathcal{L}_{recon} \quad (8)$$

where the $\alpha$ and $\beta$ are trade-off parameters.

Network training was conducted using the TensorFlow framework in an NVIDIA GPU hardware environment. The Adam optimizer was employed for parameter updates, with an initial learning rate of $5\times10^{-5}$, a weight decay of $10^{-3}$, and a batch size of 16 [59]. This configuration was maintained throughout the total 200 training epochs to ensure steady convergence of the physics-informed loss function without dynamic scheduling.

**Simulation dataset generation**

To generate gradient porosity bone models, we applied threshold-based segmentation to 31 Micro-CT scans of trabecular bone tissue with a minimum pore size of 75 μm. By varying the binarization threshold, we systematically created models with porosities ranging from 35% to 95% at seven discrete levels (35%, 45%, 55%, 65%, 75%, 85%, and 95%), with each threshold yielding distinct pore size distributions and trabecular connectivity patterns. For each porosity level, 40 independent samples were generated by randomly cropping different regions from the 31 original CT scans, resulting in a total of 280 samples. Photoacoustic wave propagation simulations were conducted using the open-source k-Wave MATLAB toolbox [45]. The computational domain was discretized into a two-dimensional grid of 1024 points along both x and y directions, with a spatial step size of 37.5 μm, yielding a domain size of 38.4 mm in both dimensions. A perfectly matched layer (PML) with 20-point thickness (0.75 mm) was set around the computational domain to absorb outgoing waves. The initial pressure distribution was uniformly assigned within the bone tissue. A 128-element linear transducer array with an inter-element spacing of 112.5 μm (3 grid points) was positioned 10 mm above the bone sample to detect the propagating acoustic waves. The temporal sampling interval was 1.3125 ns. Although each simulation ran for a total of 55 μs, a specific temporal window of 4.73-26.22 μs was extracted from each waveform for subsequent analysis. This windowing ensured the capture of essential poroelastic wave components while effectively excluding the initial propagation delay through the coupling medium and late-stage reflections.

**Ex vivo dataset generation**

Bovine femoral epiphyseal trabecular bone was used as an ex vivo surrogate for human bone tissue. Fresh bovine femoral epiphyses were sectioned into 2.5-mm-thick slices using a bone saw, then machined into standardized cylindrical specimens (20 mm diameter) using a precision grinder. All specimens were immediately preserved in 4% paraformaldehyde (PFA) fixation solution. To construct a gradient model representing different stages of bone degradation, specimens underwent differential decalcification treatment using ethylenediaminetetraacetic acid (EDTA) solution [60]. By controlling decalcification duration, 23 specimens with varying degrees of mineralization were prepared to simulate the progressive pathological process of bone degenerative diseases. All specimens were scanned using a micro-computed tomography (micro-CT) system (VENUS VNC-102, PINGSENG Healthcare, Kunshan, China) with the following parameters: voxel size of 26 $\mu m$, tube voltage of 90 kV, and tube current of 0.07 mA. Quantitative analysis of reconstructed images was performed using MATLAB. Morphological dilation operations were employed to refine trabecular thickness measurements, and corresponding porosity parameters were calculated as the gold standard for bone microstructure assessment. The 23 specimens were classified into four categories: healthy group (n=5, porosity <60%), early-stage group (n=6, porosity 60-70%), mid-stage group (n=6, porosity 70-80%), and advanced group (n=6, porosity >80%).

The photoacoustic detection setup is illustrated in Fig. 4a. A tunable optical parametric oscillator laser (Radiant OPO, OPOTEK, USA) served as the excitation source, operating at 700 nm wavelength with a pulse repetition rate of 10 Hz and pulse width of 5 ns. The laser beam was expanded through a beam expander to create a uniform illumination area of 20 mm diameter on the sample surface, with a stable laser fluence of 18.2 mJ/cm². Photoacoustic signal acquisition was performed using a 128-element linear array ultrasound transducer (L11-5v, Verasonics Inc., Kirkland, WA, USA) with a bandwidth of 5-11 MHz and element pitch of 0.3 mm. Signal reception and digitization were accomplished through a commercial ultrasound scanning platform (Vantage, Verasonics, WA, USA) with a sampling rate of 31.25 MHz. All measurements were conducted in a deionized water environment. Prior to measurement, bone specimens were subjected to vacuum degassing treatment in deionized water for 12 h to remove air bubbles trapped within the trabecular structure, ensuring optimal acoustic coupling [61]. Specimens were vertically mounted on a rotation stage via a custom fixture, with the circular cross-section perpendicular to the rotation axis. The laser beam was horizontally directed to illuminate the circular lateral surface of the specimen, while the ultrasound transducer was positioned vertically above the specimen. The transducer face was maintained at 15 mm distance from the specimen's top surface, with the linear array center aligned with the specimen's central axis. Each specimen underwent eight rotational measurements around the vertical axis at 45° intervals. At each angular position, photoacoustic signals were acquired and averaged 10 times to enhance signal-to-noise ratio. Raw radio frequency (RF) signals were digitally acquired and stored for subsequent analysis. To ensure a complete capture of the acoustic response across the 20-mm specimen diameter, a temporal window of 8-30 $\mu s$ was defined for signal processing. The starting point of 8 $\mu s$ was chosen to encompass the initial wavefront arrival after the 15-mm water-coupling delay, while the 30-$\mu s$ cutoff provided sufficient duration to record the late-arriving scattering signals and slow-wave components characteristic of the poroelastic bone matrix.

**ACKNOWLEDGEMENTS**

This project was supported by the National Natural Science Foundation of China (12034015, 62088101), the Shanghai Municipal Science and Technology Major Project (2021SHZDZX0100), and the Natural Science Foundation of Shandong Province (ZR2024QA049).

**AUTHOR CONTRIBUTIONS**

Shoukun Lyu and the corresponding authors (Ya Gao and Qian Cheng) conceived the idea of the research framework. Shoukun Lyu designed, optimized, and utilized the machine learning models. Haohan Sun and Shibo Nie assisted in the experimental data collection. Weiya Xie participated in the numerical simulations. Ying Gu participated in the model construction. Shiying Wu contributed to the data analysis. Ya Gao and Qian Cheng supervised the complete process. The manuscript was written with the contributions of all authors. All authors have approved the final version of the manuscript.